# Thermally Driven Long Range Magnon Spin Currents in Yttrium Iron Garnet due to Intrinsic Spin Seebeck Effect

Brandon L. Giles[1], Zihao Yang[2], John S. Jamison[1], Juan M. Gomez-Perez[4], Saül Vélez[4], Luis E. Hueso[4,5], Fèlix Casanova[4,5], and Roberto C. Myers[1-3]

[1]Department of Materials Science and Engineering, The Ohio State University, Columbus, OH, 43210, USA

[2]Department of Electrical and Computer Engineering, The Ohio State University, Columbus, OH, 43210, USA

[3] Department of Physics, The Ohio State University, Columbus, OH, USA

[4]CIC nanoGUNE, 20018 Donostia-San Sebastian, Basque Country, Spain

[5]IKERBASQUE, Basque Foundation for Science, 40813 Bilbao, Basque Country, Spain

Email: myers.1079@osu.edu , Web site: http://myersgroup.engineering.osu.edu

**The longitudinal spin Seebeck effect refers to the generation of a spin current when heat flows across a normal metal/magnetic insulator interface. Until recently, most explanations of the spin Seebeck effect use the interfacial temperature difference as the conversion mechanism between heat and spin fluxes. However, recent theoretical and experimental works claim that a magnon spin current is generated in the bulk of a magnetic insulator even in the absence of an interface. This is the so-called intrinsic spin Seebeck effect. Here, by utilizing a non-local spin Seebeck geometry, we provide additional evidence that the total magnon spin current in the ferrimagnetic insulator yttrium iron garnet (YIG) actually contains two distinct terms: one proportional to the gradient in the magnon chemical potential (pure magnon spin diffusion), and a second proportional to the gradient in magnon**

temperature ($\nabla T_m$). We observe two characteristic decay lengths for magnon spin currents in YIG with distinct temperature dependences: a temperature independent decay length of ~ 10 $\mu$m consistent with earlier measurements of pure ($\nabla T_m = 0$) magnon spin diffusion, and a longer decay length ranging from about 20 $\mu$m around 250 K and exceeding 80 $\mu$m at 10 K. The coupled spin-heat transport processes are modeled using a finite element method revealing that the longer range magnon spin current is attributable to the intrinsic spin Seebeck effect ($\nabla T_m \neq 0$), whose length scale increases at lower temperatures in agreement with our experimental data.

Recently, significant efforts have focused on understanding magnon spin diffusion arising from the spin Seebeck effect [1,2]. In particular, the effective magnon spin diffusion length in YIG has been experimentally measured using many different methods, including the systematic variation of YIG sample thickness to observe the effect on the longitudinal spin Seebeck signal [3–5], and by the use of a non-local geometry to directly measure the magnon spin diffusion length of electrically and thermally excited magnons [6–8]. Both methods demonstrated that the magnon spin diffusion length in YIG is only minimally dependent on film thickness and also that the magnon spin diffusion length is around 10 $\mu$m at low temperatures. However, the studies report contradictory results near room temperature. The thickness dependence study carried out by Kehlberger *et. al.* [3] found that the magnon spin diffusion length gradually decreases from 10 to 1 $\mu$m as the temperature is increased to room temperature, while the non-local measurement carried out by Cornelissen *et. al.* [7] found that the magnon spin diffusion length is only very slightly dependent on temperature. These discrepancies might be expected due to variation in the temperature profile between experiments with different sample sizes and geometries, and the variation in the relative impact of the intrinsic (bulk) spin Seebeck effect. The need to include these bulk temperature gradient driven magnon currents to fully explain room temperature nonlocal spin transport in thin film YIG has recently been discussed in detail in Ref. [8].

In this Rapid Communication, we further demonstrate the central role of the intrinsic spin Seebeck effect in the generation of long-range spin signals in bulk YIG that emerge at low temperatures. For this purpose, we carry out two independent experiments to measure diffusive magnon spin currents in bulk single crystal YIG as a function of temperature using the nonlocal opto-thermal [9] and the nonlocal electro-thermal [6] techniques. For both measurements, magnons carrying spin angular momentum are thermally excited beneath a Pt injector resulting in a measureable voltage induced in an electrically isolated Pt spin detector. In both the opto-thermal and electro-thermal measurements, two independent magnon spin current decay lengths are observed. The shorter decay length ~10 $\mu$m is roughly temperature independent and in agreement with Cornelissen *et al*. [6]. In addition to this shorter decay length, we also identify a longer range magnon spin decay length at lower temperatures that reaches values in excess of 80 $\mu$m at 10 K. The longer magnon spin decay length originates from magnons generated by heat flow within the bulk YIG itself, and represents the intrinsic spin Seebeck effect. Finite element modeling (FEM) is used to solve coupled spin-heat transport equations in YIG that describe both the pure magnon spin diffusion that is driven by a gradient in the magnon chemical potential, $\nabla \mu_m$, and also the magnon spin current that is driven by a thermal gradient in the YIG itself, $\nabla T_m$.

Microscope images of typical devices used for opto-thermal measurements and electro-thermal measurements are shown in Fig. 1(a) and Fig. 1(c). The opto-thermal device consists of 10 nm of Pt that was sputter deposited onto a 500 $\mu$m <100> single crystal YIG that was purchased commercially from Princeton Scientific. Standard lithography techniques were used to pattern the Pt into a 50×50 $\mu$m detection pad surrounded by electrically isolated 5×5 $\mu$m injector pads with 3 μm between them. The electro-thermal device consists of 5 nm of Pt that was sputter deposited onto a 500 μm <100> single crystal YIG from the same wafer. Each electro-thermal device was fabricated *via* high-resolution e-beam lithography using a negative resist and Ar-ion milling to pattern one Pt injector and two Pt detectors (width $W = 2.5$ μm and length $L = 500$ μm). Injector-detector distances range from 12 to 100 μm.

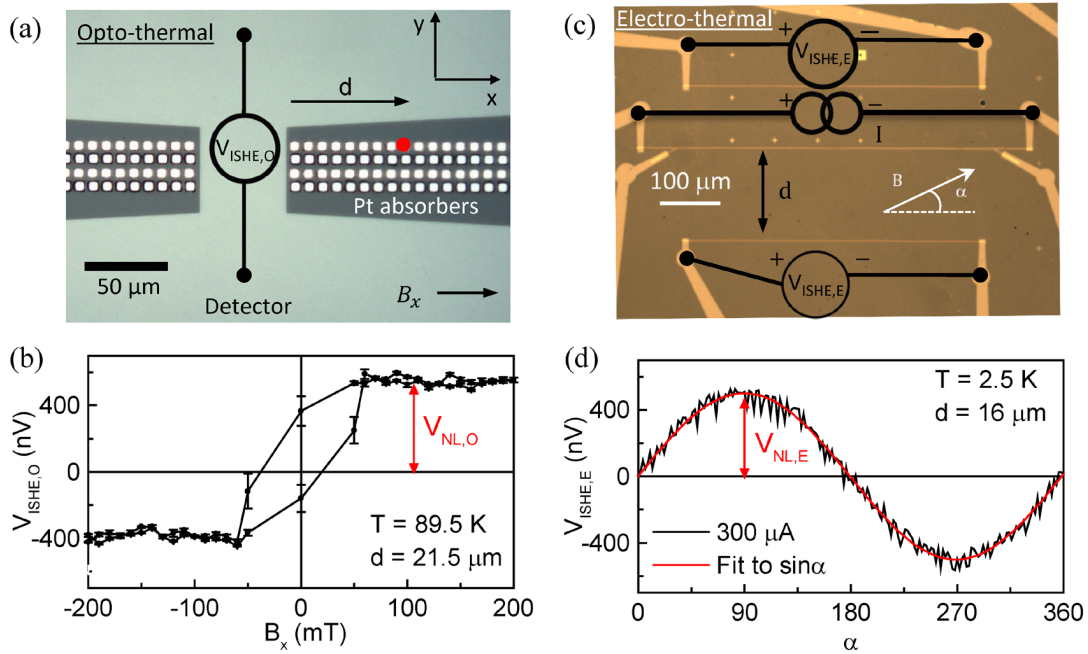

FIG 1. Optical images of the devices used in the opto-thermal and electro-thermal measurements. (a) In the opto-thermal measurement, a laser is used to thermally excite magnons in YIG beneath a Pt injector. The magnons diffuse laterally and are converted into a measureable voltage in the Pt detector. (b) A typical hysteresis loop showing the measured voltage as a function of magnetic field. $V_{NL,O}$ is defined as the magnitude of the hysteresis loop. (c) In the electro-thermal measurement, current flowing through the injector causes resistive heating, resulting in the excitation of magnons into YIG. The non-equilibrium magnons produced diffuse to the region beneath a non-local Pt detector, where can be detected due to the inverse spin Hall voltage induced. (d) The measured voltage depends sinusoidally on the angle $\alpha$ of the applied in-plane magnetic field. The maximum detected voltage is defined as $V_{NL,E}$. $d$ represents the distance the magnons have diffused from the injection to the detection site.

In the opto-thermal experiment a diffraction-limited 980-nm-wavelength laser is used to thermally excite magnons beneath a Pt injector whose center is located at a distance $d$ from the closest edge of the Pt detector. The experiments were carried out in a Montana Instruments C2 cryostat at temperatures between 4 and 300 K. The laser is modulated at 10 Hz and a lock-in amplifier referenced to the laser chopping frequency is used to measure the inverse spin Hall effect voltage, defined as $V_{ISHE,O}$, across the detector. An in-plane magnetic field is applied along the x axis and is swept from -200 mT to 200 mT while $V_{ISHE,O}$ is continuously recorded. A representative hysteresis loop taken at 89.5 K and for $d = 21$ $\mu$m is shown in Fig. 1(b). The detector signal proportional to nonlocal magnon spin diffusion, defined as $V_{NL,O}$, is obtained by taking half the difference between saturated $V_{ISHE,O}$ values at positive and negative fields, i.e. the height of the hysteresis loop. For the electro-thermal experiment, magnetotransport measurements were carried out using a Keithley 6221 sourcemeter and a 2182A nanovoltmeter operating in delta mode. In contrast to the standard current-reversal method, where one obtains information about the electrically excited magnons in devices of this kind [10], here a dc-pulsed method is used where the applied current is continuously switched on and off at a frequency of 20 Hz. This measurement provides equivalent information as the second harmonic in ac lock-in type measurements [11], i.e., it provides information about the thermally excited magnons. A current of $I = 300$ $\mu$A was applied to the injector. The experiments were carried out in a liquid-He cryostat at temperatures between 2.5 and 10 K. A magnetic field of H = 1 T was applied in the plane of the sample and rotated (defined by the angle $\alpha$) while the resulting voltage $V_{ISHE,E}$ was measured in one of the detectors. Fig. 1(d) shows a representative measurement. The signal obtained is proportional to sin $\alpha$, which is indicative of the diffusive magnon spin current [12]. The magnitude of the signal is defined as $V_{NL,E}$ [see Fig. 1(b)].

The magnon spin current decays exponentially with $d$ [13]. Therefore, the $V_{NL}$ measured in our devices is given by

$$V_{NL} = A_o e^{-\frac{\lambda_S^*}{d}}, \quad (1)$$

where $A_0$ is a pre-factor that is independent of $d$ and $\lambda_S^*$, is the effective magnon spin diffusion length. The experimental data obtained for both the opto-thermal and the electro-thermal magnon spin excitation are shown in Fig. 2 and analyzed using Eq. (1). At high temperatures, the data fits very well to a single exponential as expected. Surprisingly, at low temperatures, the fit analysis reveals that there must actually be two different decay lengths. For instance, for the opto-thermal case, it is observed that the quality of the fit rapidly decreases below a correlation coefficient of $r^2=0.985$ when the distances considered range from the smallest measured (5.5 $\mu$m) to greater than 37.5 $\mu$m. This indicates that the application of the spin decay model is only appropriate up to 37.5 $\mu$m. If distances greater than 37.5 $\mu$m are considered and the data is fit to Eq. (1), a lower $r^2$ factor is obtained, indicating a low quality fit. This observation inspires us to separate the $V_{NL,O}$ data into two distinct regions defined as the $\lambda_1$ and $\lambda_2$ regions [see Fig. 2(a)]. Equation (1) is fit to each individual region. The effective magnon spin diffusion length $\lambda_S^*$ is extracted for each region separately and plotted in Fig. 3. The same analysis was performed for the electro-thermal measurements and the existence of two different decay lengths was confirmed (See Fig. 2(b)).

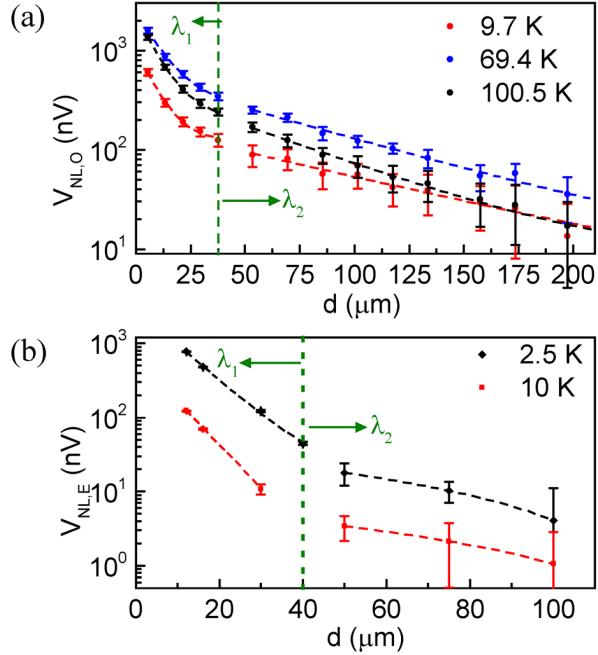

FIG 2. (a) $V_{NL,O}$ as a function of $d$ with the measurement shown at different temperatures. The measurement results are divided into two regions defined as $\lambda_1$ and $\lambda_2$. Dotted lines represent single exponential fits of the data to Eq. (1) in each region. The decay in $\lambda_1$ is shorter, while it appears to be much longer in $\lambda_2$. (b) $V_{NL,E}$ as a function of $d$ with the measurement shown at multiple temperatures. Dividing the data also into the $\lambda_1$ and $\lambda_2$ regions confirms the existence of the two different characteristic decay lengths. Dashed lines are fits to Eq. (1) in each region.

Fig. 3 shows the extracted values of the magnon spin diffusion lengths in each of the two regions as a function of temperature for both the opto-thermal and electro-thermal measurements. At low temperature, both measurements indicate an effective spin diffusion length of about 10 μm in the $\lambda_1$ region, which is in excellent agreement with previously reported values and temperature dependence of the magnon spin diffusion length [7]. Note that in the earlier opto-thermal study [9] the data indicated only a single exponential decay, which was interpreted as the spin diffusion length. In the opto-thermal measurements reported here, the improved signal to noise ratio of the experiment reveals the double exponential character of the spin decay profile. The current data can still be fitted to a single exponential decay at 23 K of 47 μm, consistent with the earlier report, however the improved data set in the current study demonstrates that a double exponential decay fit is far better quality.

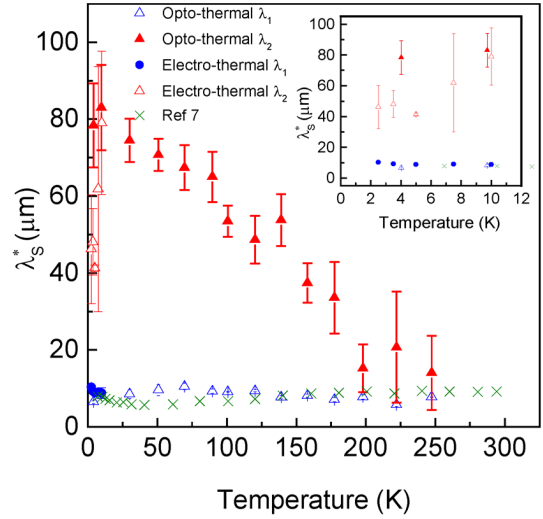

FIG 3. The extracted decay parameters $\lambda_S^*$ from the $\lambda_1$ and $\lambda_2$ regions as a function of temperature and for both experiments. $\lambda_S^*$ values reported in Ref. 7 are included for comparison. Inset: zoomed view of low temperature data.

A larger $\lambda_S^*$ in the $\lambda_2$ region is observed in both the opto-thermal and electro-thermal measurements. At temperatures above 10 K in the electro-thermal measurement, the non-local signal magnitude strongly decreased and could not be measured at enough values of d in order to make a meaningful exponential fit to extract $\lambda_S^*$ in the $\lambda_2$ region. The effective magnon spin diffusion length in the $\lambda_2$ region is approximately one order of magnitude larger than in the $\lambda_1$ region at low temperatures and decreases monotonically with increasing temperature. The maximum value of 83.03 $\mu$m occurs at 9.72 K and the minimum value of 14.05 $\mu$m at 247.5 K. A zoom of the data at low T is shown in the inset to Fig. 3. In the electro-thermal measurements, the maximum value of $\lambda_2$ is not at the lowest temperature, but at ~10 K in agreement with the optothermal measurements. This is consistent with the origin of $\lambda_2$ as from intrinsic SSE associated with the temperature profile in YIG since as T approaches 0 K, thermal conductivity becomes negligible.

To justify the existence of the long range spin current persisting well beyond the intrinsic magnon spin diffusion length, the measurements are compared to a simulation of the diffusive transport of thermally generated magnons, which is obtained using three dimensional (3D) finite element modeling (FEM). The simulation is solved using COMSOL Multiphysics and is based on the spin and heat transport formalism that is developed in [14,15].

In the simulation, the length scale of the inelastic phonon and magnon scattering is assumed to be small, implying that the phonon temperature, $T_p$, is equal to the magnon temperature $T_m$ over the lengths of interest. In addition, the simulation neglects the spin Peltier effect. Thus, the spin and heat transport equations are only partially coupled.

The simplified spin transport equation that is used to model the magnon spin current within YIG is

$$\sigma \nabla^2 \mu + \varsigma \nabla^2 T = g\mu \tag{2}$$

and the Pt/YIG interfacial boundary condition states

$$j_{m,z} = \sigma \nabla \mu_z + \varsigma \nabla T_z = G_S \mu \tag{3}$$

where $j_{m,z}$ is the simulated spin current perpendicular to the Pt/YIG interface, $\sigma$ is the spin conductivity in the YIG, $\mu$ is the magnon chemical potential, $\varsigma$ is the intrinsic spin Seebeck coefficient, $g$ describes the magnon relaxation, $T = T_p \sim T_m$ is the temperature in YIG, $G_S$ is the interfacial magnon spin conductance, and $\nabla\mu_z$ and $\nabla T_z$ represent the gradient of the magnon chemical potential and temperature along the direction perpendicular to the Pt/YIG interface, respectively.

We first solve for the temperature profile in a simulated Pt/YIG system using the parameters listed in Table I. The geometry of the model is the same as the experimental geometry of the opto-thermal measurement including the Pt absorbers. As previously stated, $d$ is defined as the distance from the edge of the Pt detector to the center of the (simulated) laser heat source at the center of the absorber.

Table I – Parameters used in the 3D FEM modeling. $\sigma$ and $G_S$ are calculated based on data reported in [15]. $\kappa_{YIG}$ is taken from [19] and $\kappa_{Pt}$ is from [20].

| $T$(K) | $\sigma$(J/mV) | $G_S$(S/m$^2$) | $\kappa_{YIG}$ (W/mK) | $\kappa_{Pt}$ (W/mK) |
|---|---|---|---|---|
| 10 | $3.10 \times 10^{-8}$ | $5.84 \times 10^{10}$ | 60.00 | 1214.98 |
| 70 | $8.32 \times 10^{-8}$ | $1.08 \times 10^{12}$ | 37.59 | 91.82 |
| 175 | $1.32 \times 10^{-7}$ | $4.27 \times 10^{12}$ | 11.41 | 75.56 |
| 300 | $1.73 \times 10^{-7}$ | $9.60 \times 10^{12}$ | 6.92 | 73.01 |

The decay profile for the interfacial spin current $j_{m,z}$ is obtained by using the calculated temperature profile as an input in Eq. (3). We report the total interfacial spin current that reaches the detector $j_{m,z}$ by evaluating the surface integral $\iint j_{m,z}(x,y)dA$ beneath the detector. The decay profile is calculated as a function of simulated laser position, at multiple different temperatures, ranging from 5 – 300 K. The values of the physical parameters used in the model are recorded in Table I.

From Eq. (3) one can see that $\boldsymbol{j}_{m,z}$ can be broken up into two components $j_{m,z}^{\nabla\mu}$, which is a component that is proportional to the interfacial gradient of the magnon chemical potential, and $j_{m,z}^{\nabla T}$, which is a component that is proportional to the interfacial gradient of the magnon temperature. The decomposition of the simulated spin current at the detector is shown in Fig. 4(a), which depicts a representative plot of the total $\boldsymbol{j}_{m,z}$ as a function of $d$ at 70 K, as well as the components $j_{m,z}^{\nabla\mu}$ and $j_{m,z}^{\nabla T}$. By analyzing the decay lengths of these individual components of $\boldsymbol{j}_{m,z}$ separately, it is possible to qualitatively understand the existence of the experimentally observed short and long range decay lengths.

As shown in Fig. 4(a), the component of $\boldsymbol{j}_{m,z}$ that is proportional to $\nabla\mu$ decays much more rapidly than the component of $\boldsymbol{j}_{m,z}$ that is proportional to $\nabla T$. This indicates that the total spin current that reaches the Pt detector should consist of a shorter decay component and a longer decay component. We hypothesize that the driving force of the shorter range component is the gradient of the magnon chemical potential, $\nabla\mu$ and that the driving force of the longer range component is the gradient of the magnon temperature $\nabla T$. To verify this conjecture, the plot of the simulated $\boldsymbol{j}_{m,z}$ vs. $d$ is divided into the same $\lambda_1$ and $\lambda_2$ regions as in the opto-thermal experimental measurement (where the $\lambda_2$ region is defined as $d > 37.5$ $\mu$m). Equation (1) is fit independently to the simulated $j_{m,z}^{\nabla\mu}$ within the $\lambda_1$ region, where the shorter range driving force is expected to dominate, and to the simulated $j_{m,z}^{\nabla T}$ within the $\lambda_2$ region where the longer range driving force will be most prevalent, as shown in the representative 70 K plot in Fig. 4(a). The decay parameters of these fits, $\lambda_{\nabla\mu}^*$ and $\lambda_{\nabla T}^*$, are extracted and plotted as a function of temperature in Fig. 4(b). The intrinsic spin diffusion length, $\lambda^*_{\nabla\mu}$, is relatively constant as a function of temperature, implying that $\nabla\mu$ is responsible for the shorter range spin current observed in the $\lambda_1$ region (Fig. 3). On the other hand, the bulk generated magnon current, characterized by $\lambda^*_{\nabla T}$, decays monotonically with temperature, in agreement with the observed longer decay in the $\lambda_2$ region (Fig. 3), thus implying that $\nabla T$ is the driving force for the long range spin current. Since it is the temperature profile within YIG that determines $\lambda^*_{\nabla T}$, it will vary with the thermal boundary conditions. This explains why the long range spin current manifests in bulk YIG at low temperature [9], but not in YIG/GGG thin films [7].

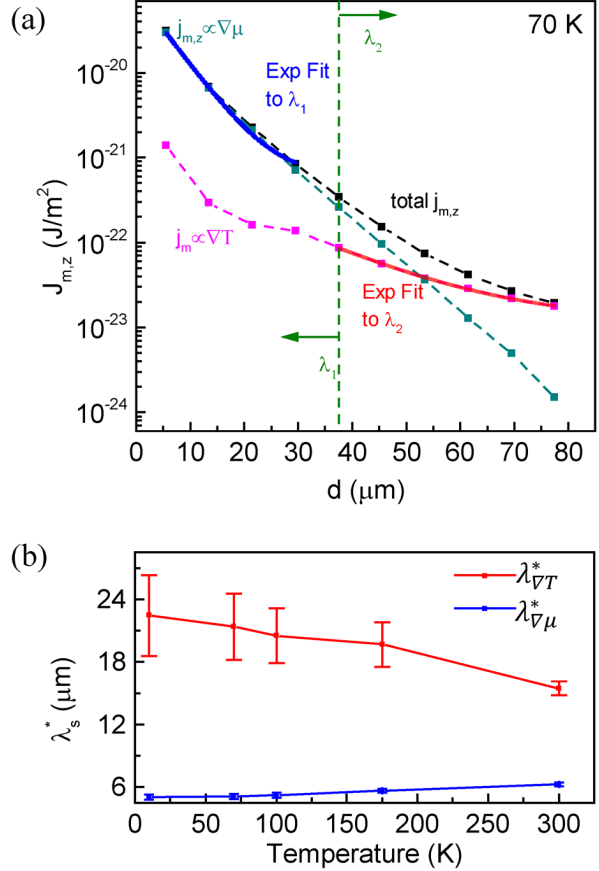

FIG 4. 3D FEM modeling simulation of the opto-thermal measurement. (a) Dashed lines represent the total spin current (black), the component of spin current proportional to $\nabla\mu$ (green) and the component of spin current proportional to $\nabla T$ (pink). Solid lines represent individual exponential fits to the corresponding component of the spin current in each of the distinct $\lambda_1$ and $\lambda_2$ regions (blue and red respectively). (b) The magnon spin diffusion lengths $\lambda^*_{\nabla\mu}$ and $\lambda^*_{\nabla T}$ extracted for each region are plotted as a function of temperature.

It should be noted that while the monotonic decay with temperature of the simulated $\lambda^*_{\nabla T}$ agrees with the measured opto-thermal and electro-thermal long range decay in the $\lambda_2$ region, the simulated magnitude of $\lambda^*_{\nabla T}$ is smaller than the one obtained experimentally. This is attributed to uncertainties in the temperature dependence of the inputs to the FEM modeling, particularly of the magnon scattering time $\tau$, which is used to calculate $\sigma_m$. At low temperatures magnon relaxation is primarily governed by magnon-phonon interactions that create or annihilate spin waves by magnetic disorder and $\tau \sim \hbar/\alpha_G k_B T$ where $\alpha_G = 10^{-4}$ [16]. This leads to calculated values of $\sigma_m$ that vary with experimental measurements by orders of magnitude [15]. Such discrepancies may be explained by recent works that attribute the primary contributors to the SSE as low-energy subthermal magnons [5,17], however an analysis of the complete temperature dependence of effective magnon scattering time based on the spectral dependence of the dominant magnons involved in SSE is outside the scope of this work. Another source of uncertainty in the simulations is the role of spin sinking into the Pt absorbers (present in the opto-thermal measurements) on the spin current decay profile. To test this, identical simulations, as described above, are carried out but with the Pt absorber pads removed. The absorbers cause a decrease in $\lambda^*_{\nabla\mu}$ of 1-2 μm, while the $\lambda^*_{\nabla T}$ shows no significant change within the uncertainty. During the review of this paper, we became aware of a related paper discussing the role of intrinsic spin Seebeck in the nonlocal spin currents decay profile [18].

In conclusion, opto-thermal and electro-thermal measurements independently demonstrate the existence of a longer range magnon spin current at low temperatures persisting well beyond the intrinsic spin diffusion length. By representing the total magnon spin current by its individual components, one of which is proportional to the gradient in magnon chemical potential and the other of which is proportional to the gradient in magnon temperature, the driving force of the longer range magnon spin diffusion can be attributed to the gradient in magnon temperature, i.e. the intrinsic spin Seebeck effect.

The authors thank Bart van Wees, Ludo Cornelissen, Yaroslav Tserkovnyak and Benedetta Flebus for valuable discussions. This work was primarily supported by the Army Research Office MURI

W911NF-14-1-0016. J.J. acknowledges the Center for Emergent Materials at The Ohio State University, an NSF MRSEC (Award Number DMR-1420451), for providing partial funding for this research. The work at CIC nanoGUNE was supported by the Spanish MINECO (Project No. MAT2015-65159-R) and by the Regional Council of Gipuzkoa (Project No. 100/16). J.M.G.-P. thanks the Spanish MINECO for a Ph.D. fellowship (Grant No. BES-2016-077301).

[1] K. Uchida, M. Ishida, T. Kikkawa, A. Kirihara, T. Murakami, and E. Saitoh, J. Phys. Condens. Matter **26**, 343202 (2014).
[2] A. Prakash, J. Brangham, F. Yang, and J. P. Heremans, Phys. Rev. B **94**, 014427 (2016).
[3] A. Kehlberger, U. Ritzmann, D. Hinzke, E.-J. Guo, J. Cramer, G. Jakob, M. C. Onbasli, D. H. Kim, C. A. Ross, M. B. Jungfleisch, B. Hillebrands, U. Nowak, and M. Kläui, Phys. Rev. Lett. **115**, 096602 (2015).
[4] E.-J. Guo, J. Cramer, A. Kehlberger, C. A. Ferguson, D. A. MacLaren, G. Jakob, and M. Kläui, Phys. Rev. X **6**, 031012 (2016).
[5] T. Kikkawa, K. Uchida, S. Daimon, Z. Qiu, Y. Shiomi, and E. Saitoh, Phys. Rev. B **92**, 064413 (2015).
[6] L. J. Cornelissen, J. Liu, R. A. Duine, J. B. Youssef, and B. J. van Wees, Nat Phys **11**, 1022 (2015).
[7] L. J. Cornelissen, J. Shan, and B. J. van Wees, Phys. Rev. B **94**, 180402 (2016).
[8] J. Shan, L. J. Cornelissen, N. Vlietstra, J. Ben Youssef, T. Kuschel, R. A. Duine, and B. J. van Wees, Phys. Rev. B **94**, 174437 (2016).
[9] B. L. Giles, Z. Yang, J. S. Jamison, and R. C. Myers, Phys. Rev. B **92**, 224415 (2015).
[10] S. Vélez, A. Bedoya-Pinto, W. Yan, L. E. Hueso, and F. Casanova, Phys. Rev. B **94**, 174405 (2016).
[11] F. L. Bakker, A. Slachter, J.-P. Adam, and B. J. van Wees, Phys. Rev. Lett. **105**, 136601 (2010).
[12] S. R. Boona, R. C. Myers, and J. P. Heremans, Energy Environ. Sci. **7**, 885 (2014).
[13] T. Valet and A. Fert, Phys. Rev. B **48**, 7099 (1993).
[14] B. Flebus, S. A. Bender, Y. Tserkovnyak, and R. A. Duine, Phys. Rev. Lett. **116**, 117201 (2016).
[15] L. J. Cornelissen, K. J. H. Peters, G. E. W. Bauer, R. A. Duine, and B. J. van Wees, Phys. Rev. B **94**, 014412 (2016).
[16] S. Hoffman, K. Sato, and Y. Tserkovnyak, Phys. Rev. B **88**, 064408 (2013).
[17] I. Diniz and A. T. Costa, New J. Phys. **18**, 052002 (2016).
[18] J. Shan, L. J. Cornelissen, J. Liu, J. B. Youssef, L. Liang, and B. J. van Wees, ArXiv170906321 Cond-Mat (2017).
[19] S. R. Boona and J. P. Heremans, Phys. Rev. B **90**, 064421 (2014).
[20] J. E. Jensen, W. A. Tuttle, H. Brechnam, and A. G. Prodell, *Brookhaven National Laboratory Selected Cryogenic Data Notebook* (Brookhaven National Laboratory, New York, 1980).